\documentclass[letter,10pt]{article}
\usepackage[utf8]{inputenc}
\pdfoutput=1
\usepackage{graphicx}
\usepackage{natbib}
\usepackage{hyperref}
\usepackage{courier}
\usepackage[margin=2cm]{geometry}
\usepackage[table,svgnames]{xcolor}
\usepackage{multirow}
\usepackage{tabu}
\usepackage{float}
\usepackage{listings}
\usepackage{xspace}
\usepackage{makecell}
\usepackage[toc,page]{appendix}
\definecolor{HeaderColor}{rgb}{0.8, 0.8, 0.95}
\usepackage[font={small}]{caption}
\usepackage[T1]{fontenc}
\usepackage[utf8]{inputenc}
\usepackage{authblk}

\author[1,2,3]{Dimitri Yatsenko\thanks{dvyatsen@bcm.edu}}
\author[1,2,3]{Edgar Y. Walker}
\author[1,2,3]{Andreas S. Tolias}
\affil[1]{Center for Neuroscience and Artificial Intelligence, Baylor College of Medicine, Houston, Texas, USA}
\affil[2]{Department of Neuroscience, Baylor College of Medicine, Houston, Texas, USA}
\affil[3]{Vathes LLC, Houston, Texas, USA}

\newcommand{\datajoint}{DataJoint\xspace}
\date{\today\\Revision 1.0}

\graphicspath{{./figures/}}

\lstdefinelanguage[]{SSQL}[]{SQL}{
  morecomment=[l][\color{Sienna}\itshape\small]{\#},
  morekeywords={COMMENT, REFERENCES, unsigned}}

\lstdefinelanguage{dj}{
  keywords={int, smallint, char, varchar, enum, unsigned, date, year, decimal, insert, delete, update, double},
  keywordstyle=\color{blue},
  keywords=[2]{boolean, string, number, objectid},
  keywordstyle=[2]\color{green}\bfseries,
  identifierstyle=\color{Black},
  sensitive=true,
  morecomment=[l][\color{Teal}\bfseries]{::},
  morecomment=[l][\color{Sienna}\itshape\small]{\#},
  stringstyle=\color{Indigo},
  morestring=[b]',
  morestring=[b]"
}

\newfloat{lstfloat}{htbp}{lop}

\title{\datajoint: A Simpler Relational Data Model}

\begin{document}

\maketitle
\begin{abstract}
The relational data model offers unrivaled rigor and precision in defining data structure and querying complex data. 
Yet the use of relational databases in scientific data pipelines is limited due to their perceived unwieldiness.
We propose a simplified and conceptually refined relational data model named \datajoint. 
The model includes a language for schema definition, a language for data queries, and diagramming notation for visualizing entities and relationships among them.  
The model adheres to the principle of \emph{entity normalization}, which requires that all data --- both stored and derived --- must be represented by well-formed entity sets.
\datajoint's data query language is an algebra on entity sets with five operators that provide matching capabilities to those of other relational query languages with greater clarity due to entity normalization. 
Practical implementations of \datajoint have been adopted in neuroscience labs for fluent interaction with scientific data pipelines.
\end{abstract}
\tableofcontents 

\twocolumn

\section{The Relational Data Model}
\subsection{Core concepts and terminology}
The relational data model \citep{codd_relational_1970} provides the most rigorous approach to structured data storage and the most precise approach to querying data.  
The model is defined by the principles of data representation, domain constraints, uniqueness constraints, referential constraints, and declarative queries as summarised in Table \ref{tab:core}.

\tabulinesep=6pt
\begin{table*}[ht]
\begin{tabu}{|X|}
\hline
{\bf Data representation.} Data are represented and manipulated in the form of \emph{relations}. 
A relation is a \emph{set} (\emph{i.e.}\ an unordered collection) of \emph{tuples} of values for each of the respective named \emph{attributes} of the relation.
\emph{Base relations} represent stored data while \emph{derived relations} are formed from base relations through query expressions.
A collection of base relations with their attributes, domain constraints, uniqueness constraints, and referential constraints is called a \emph{schema}.

\\
{\bf Domain constraints.} Attribute values are drawn from corresponding attribute \emph{domains}, \emph{i.e.}\ predefined sets of values.
Attribute domains may not include relations, keeping data model flat, free of nested structures.

\\
{\bf Uniqueness constraints.} Tuples within relations are addressed by values of their attributes.
To identify and relate data elements, \emph{uniqueness constraints} are imposed on subsets of attributes, then referred to as \emph{keys}. 
One key in a relation is designated as the \emph{primary key} used for referencing its elements.

\\
{\bf Referential constraints.} Associations among data are established by means of \emph{referential constraints} with the help of \emph{foreign keys}. 
A referential constraint on relation {\tt A} referencing relation {\tt B} allows only those tuples in {\tt A} whose foreign key attributes match the key attributes of a tuple in {\tt B}. 

\\
{\bf Declarative queries.} Data \emph{queries} are formulated through declarative, as opposed to imperative, specifications of sought results. 
This means that \emph{query expressions} convey the logic for the result rather than the procedure for obtaining it. 
Formal languages for query expressions include \emph{relational algebra}, \emph{relational calculus}, and SQL {\tt SELECT} statements.
\\
\hline
\end{tabu}
\caption{Core principles of the relational data model.}
\label{tab:core}
\end{table*}

Popular implementations of the relational data model rely on the Structured Query Language (SQL).
SQL comprises distinct sublanguages for schema definition, data manipulation, and data queries.
SQL thoroughly dominates in the space of relational databases and is often conflated with the relational data model in casual discourse. 

Various terminologies are used to describe related concepts from the relational data model (Table \ref{tab:terms}).
Relations are often visualized as \emph{tables} with attributes corresponding to \emph{columns} and tuples corresponding to \emph{rows}.  
In particular, SQL uses the terms \emph{table}, \emph{column}, and \emph{row}.  

\tabulinesep=6pt
\begin{table*}[ht]
   \rowcolors{1}{white}{gray!20}
   \begin{tabu}{|X[1,c,p]| X[1,c]| X[1,c]| X[1,c]|}
   \hline
   \rowcolor{HeaderColor}
   {\bf Relational} & {\bf ERM} & {\bf SQL} & {\bf \datajoint}  \\
   \cellcolor{white} & entity set & \cellcolor{white} & \cellcolor{white} \\
   \multirow{-2}{*}{relation}  & relationship set  & \multirow{-2}{*}{table}  &  \multirow{-2}{*}{entity set} \\
   tuple       & entity           & row    & entity \\
   domain      & value set        & data type & data type \\
   attribute   & attribute        & column {\em or} field    & attribute \\
   attribute value & attribute value  & field value & attribute value \\
   primary key & primary key & primary key & primary key \\
   foreign key & foreign key & foreign key & foreign key {\em or} dependency \\
   schema      & schema      &  schema  &  schema \\
   relational expression \par {\em or} derived relation &  data query & {\tt SELECT} statement & query expression \\
   \hline
   \end{tabu}
\caption{Corresponding terms used in variants of relational models.}
\label{tab:terms}
\end{table*}

\subsection{Conceptual clarification}
The relational model is abstract and semantically unconstrained, providing few guidelines for translating problems into database schemas or for forming valid queries. 
Learning to design databases and to compose valid queries in practical problems requires some form of \emph{conceptual clarification}: a system of conventions and techniques for intuitive mapping from real-world rules to relational concepts.

A set of formal rules known as \emph{normal forms} have been devised to test whether a particular schema meets basic quality criteria that preclude redundancies in data storage and anomalies in data manipulations \citep{kent-1983-simple}.  
Relying on the mathematical concept of {\em functional dependencies} among the attributes, normal forms remain abstract without a definitive approach to conceptual modeling of real-world problems. 
When a schema does not meet the normal forms, it is considered \emph{unnormalized}; whereas redesigning schemas to meet these tests is called \emph{normalization}.
SQL does not provide mechanisms for enforcing or diagnosing normalization in its data definition language. 
Mastery of normalized schema design requires considerable training. 

For data queries, also, the relational data model provides few constraints as to what constitutes a valid and  meaningful query.
It allows unlimited freedom to compare, match, and combine attributes of relations regardless of their semantic compatibility.

New practitioners entering the field of database programming with their diverse backgrounds and mindsets may conceive of database tables as mere spreadsheets with arbitrary semantics of rows and columns.  
Extended experience with more sophisticated designs helps develop semantic mappings between real-world information motifs and database design patterns.

The lack of semantic constraint in the relational data model leaves great freedom to experienced developers with honed conceptual skills. 
But it also allows for a great diversity of incompatible approaches to schema design and data queries and lengthens the path from mediocrity to proficiency.

\subsection{The Entity-Relationship Model}\label{sec:erm}
A successful effort to clarify the relational data model was laid out by the Entity-Relationship Model (ERM) \citep{chen_entity_1976}.  
The ERM adopts the view that the world can be intuitively modeled to consist of entities and relationships between them them; it prescribes an approach to map this view into relational concepts.

The ERM provides 
\begin{enumerate}
\item a general approach for problem analysis and conceptual modeling for schema design
\item diagramming notation for schema design 
\item guidelines for composing meaningful queries
\end{enumerate}

The central concept in the ERM is that of an {\em entity set}: an unordered collection of identifiable items (entities) in the modeled world that share the same set of attributes, are distinguished from each other by the same \emph{primary key}, and can participate in the same types of relationships with other entity sets. 
In the ERM, all base relations are either \emph{entity sets} or \emph{relationship sets}; these terms effectively subsume the term \emph{relation} (See Table \ref{tab:terms}).

A \emph{relationship set} is a collection of associations linking entities from two or more entity sets. 
These associations take the form of referential constraints (foreign keys) between relationships sets and entity sets.
In addition to foreign keys, entity sets may have secondary attributes describing the relationship.

Although the ERM is mostly known for its approach to schema design, it also prescribes methods to form valid data queries: Queries rely on foreign keys to match entities through the relationships in which they participate. 

\begin{figure}
\centering
\includegraphics{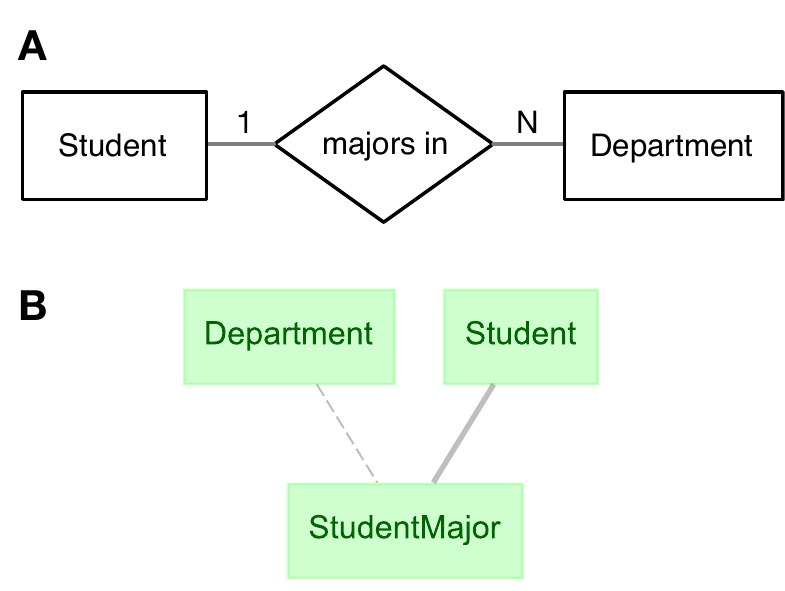}
\caption{
{\sf A:} A database schema modeling university departments, students, and their major departments in ERM diagramming notation.
{\sf B:} An equivalent \datajoint schema diagram.}
\label{fig:erm-notation}
\end{figure}

ERM diagramming notation depicts entity sets and relationship sets, turning the ERM into an effective tool for \emph{conceptual modeling} and communication between database designers, customers, and management.
For example, Figure \ref{fig:erm-notation}A depicts a simple entity-relationship diagram for a university student database with two entity sets, {\tt Student} and {\tt Department} and the relationship {\tt majors in} between them.

The ERM was proposed as a complete data model in its own right but its conceptual refinements were never translated into a practical programming language. 
Today, the ERM is best known as a diagramming technique for conceptual modeling.  
Formal courses in database design prescribe a two-phased approach: \emph{conceptual modeling} using a diagramming tool such as the ERM followed by its translation (\emph{logical  design}) into a relational database schema and SQL \citep{elmasri-2015-fundamentals, coronel-2016-database}.
Expert database designers internalize ERM precepts and commonly distinguish ``entity tables'' from ``relationship tables'' even though SQL makes no such distinctions. 

A schema definition language tailored to the ERM will provide greater clarity, making it easier to learn and to use than the more general relational model and SQL. 
ERM designs naturally produce database schemas that comply with normal forms; 
the problem of database normalization is translated into a more relatable problem of distinguishing well-defined entities in the modeled world.

For example, in the ERM, the foreign key always references the primary key of the referenced entity set.  
Therefore, an ERM-specific schema definition language can simplify the use of foreign keys.
Consider the following section of SQL table definition code defining a foreign key to the {\tt Course} table whose primary key comprises attributes ({\tt dept}, {\tt course}):
\begin{lstlisting}[language=SSQL]
dept char(6) NOT NULL COMMENT 
  "abbreviated department name, e.g. BIOL",
course int unsigned NOT NULL COMMENT 
  "course number, e.g. 1010",
FOREIGN KEY (dept, course) 
  REFERENCES Course(dept, course)
\end{lstlisting}
In an ERM-tailored language, the equivalent foreign key definition can be simply
\begin{lstlisting}[language=dj]
-> Course
\end{lstlisting}

This definition clearly and simply communicates the intent of the foreign key: reference elements of the entity set \lstinline$Course$, without extraneous detail.
The definition combines two jobs in a single step:  add the referencing attributes to the table definition and create the foreign key constraint.  
The resulting definition is not only more succinct but is also less error-prone since changes in the definition of the referenced table will correctly propagate into the referencing table. 

This syntax simplification would serve well in designs that follow ERM conventions. 
In more esoteric uses of the relational model, foreign keys may reference non-primary key attributes and may require special syntax.

The ERM's conceptual clarity comes at the expense of some loss of representational power, skirting around some of the more esoteric capabilities and complexities of the broader relational model. 
For example, ERM forgoes representations of multiple functional dependencies between overlapping sets of attributes for which the higher normal forms become applicable.
Giving up some representational power for increased conceptual clarity is a fair tradeoff in practical database design. 

\subsection{The \datajoint model}
We introduce \datajoint as a conceptual refinement of the relational data model offering a more expressive and rigorous framework for database programming.
The \datajoint model facilitates clear conceptual modeling, efficient schema design, and precise and flexible data queries. 

The model has emerged over a decade of continuous development of complex data pipelines for neuroscience experiments \citep{yatsenko-datajoint-2015}. 
\datajoint has allowed researchers with no prior knowledge of databases to collaborate effectively on common data pipelines sustaining data integrity and supporting flexible access.
\datajoint is currently implemented as client libraries in MATLAB and Python.  
These libraries work by transpiling \datajoint queries into SQL before passing them on to conventional relational database systems serving as the backend.
In this paper, we crystallize the underlying data model to guide future development.
While we focus on concepts and principles, we present the model as a complete scripting language separate from any host programming language.

\datajoint comprises 
\begin{itemize}
\item a schema definition language (Sec.\ \ref{sec:def1} and \ref{sec:def2})
\item a data manipulation language (Sec.\ \ref{sec:manip})
\item a data query language (Sec.\ \ref{sec:query})
\item diagramming notation for visualizing relationships between modeled entities (Sec.\ \ref{sec:diag}).
\end{itemize}

In \datajoint, all data exist in the form of relations representing entity sets.
As with any set, the order of elements in an entity set is never significant; elements cannot be addressed or identified by their position in the entity set.

Since entity sets are relations, which are often visualized as tables, the terms \emph{relation}, \emph{entity set}, and \emph{table} can be used interchangeably in \datajoint (See Table \ref{tab:terms}).
Individual elements of an entity set may be called \emph{entities}, \emph{entity instances}, \emph{tuples}, or \emph{rows}; 
for consistency, we will continue to refer to them as simply \emph{elements} of their respective entity sets.

The attributes of an entity set may also be called {\em columns} or {\em fields}.
In \datajoint, attributes in relations are always identified by their names and their order is never significant, from the logical point of view, although it may affect the performance of some operations.

\subsection{Entity normalization}\label{sec:norm}
\emph{Entity normalization} is a conceptual refinement of the relational data model and the central principle of the \datajoint model.
Entity normalization is the requirement that all data must exist in the form of relations that meet the criteria of \emph{well-formed entity sets}.

These criteria are
\begin{enumerate}
\item All elements of an entity set belong to the same well-defined and readily identified \emph{entity type} from the model world.
\item All attributes of an entity set are applicable directly to each of its elements, although some attribute values may be missing (set to null).  
\item All elements of an entity set must be distinguishable form each other by the same \emph{primary key}.
\item Primary key attribute values cannot be missing, \emph{i.e.}\ set to null.
\item All elements of an entity set participate in the same types of relationships with other entity sets.
\end{enumerate}

The term \emph{entity normalization} refers to the procedure of refactoring a schema design that does not meet the above criteria into one that does. 
In some cases, this may require breaking up some entity sets into multiple entity sets, causing some entities to be represented across multiple entity sets.
In other cases, this may require converting attributes into their own entity sets.

Entity normalization entails compliance with the \emph{Boyce-Codd normal form} while lacking the representational power for the applicability of more complex normal forms \citep{kent-1983-simple}.  
Thus adherence to entity normalization prevents redundancies and data manipulation anomalies that originally motivated the formulation of the classical relational normal forms. 

Adherence to entity normalization is the common thread unifying \datajoint's data definition, data manipulation, and data queries. 

\section{Schema definition}\label{sec:def1}
\subsection{Entity set definition}
\datajoint differentiates between \emph{base entity sets} and \emph{derived entity sets}: the former represent stored data and are persistent whereas the latter arise from query expressions.
\datajoint's  \emph{schema definition language} allows defining base entity sets and dependencies between them.
A \emph{schema} is a collection of base entity sets and their dependencies (See Tables \ref{tab:core} and \ref{tab:terms}).
Base entity sets represent the stored data and are identified by unique names within their schemas. 

Let's define a simple university database for use as a running example. 
The example will include student information, university courses, and enrollment information.
Variations of such university databases have been used in many papers and textbooks on relational design theory and we follow suit here.

Listing \ref{lst:uni1} begins the university database by declaring entity sets \lstinline$Student$, \lstinline$ Department$, and \lstinline$StudentMajor$ as diagrammed in Figure \ref{fig:erm-notation}B.

\begin{lstfloat*}
\begin{lstlisting}[language=dj,caption={University database schema definition (Part 1).},label={lst:uni1}]
::Student     
student_id : int unsigned   # university ID 
---
first_name      : varchar(40)
last_name       : varchar(40)
sex             : enum('F', 'M', 'U')
date_of_birth   : date
home_address    : varchar(200) # street address
home_city       : varchar(30) 
home_state      : char(2)  # two-letter abbreviation
home_zipcode    : char(10)
home_phone      : varchar(14) 

::Department 
dept : char(6)   # abbreviated department name, e.g. BIOL
---
dept_name    : varchar(200)  # full department name
dept_address : varchar(200)  # mailing address
dept_phone   : varchar(14)  

::StudentMajor
-> Student
---
-> Department
declare_date :  date  # when student declared her major  
\end{lstlisting}
\end{lstfloat*}

Each entity set begins with the line specifying the entity name as 
\begin{lstlisting}[language=dj]
::EntityName 
\end{lstlisting}

By convention, entity set names describe an individual element of the set in the singular form.  
For example, entity set {\tt Student} represents all students in our university. 

\subsubsection{Attributes and their datatypes}
Each line of the entity definition defines an entity attribute in the form
\begin{lstlisting}[language=dj]
attr_name : datatype   # comment
\end{lstlisting}
The comment is optional.

Optionally, the attribute definition may specify a default value \lstinline$v$ for the attribute:
\begin{lstlisting}[language=dj]
attr_name = v : datatype  # comment
\end{lstlisting}
The default value of \lstinline$null$ indicates that the attribute value may be altogether omitted in any element of the entity set.

A variety of datatypes may be used.  In our examples, we will use the following datatypes familiar from SQL variants.
\begin{itemize}
\item \lstinline[language=dj]$int$ -- 32-bit signed integer 
\item \lstinline[language=dj]$int unsigned$ -- 32-bit unsigned integer
\item \lstinline[language=dj]$decimal(n,m)$ -- decimal number with {\tt n} total digits and {\tt m} fractional digits
\item \lstinline[language=dj]$char(n)$, \lstinline[language=dj]$varchar(n)$ -- a string of up to {\tt n} characters
\item \lstinline[language=dj]$date$ -- calendar date
\item \lstinline[language=dj]$year$ -- calendar year 
\item \lstinline[language=dj]$enum('one', 'two' , 'three')$ -- element from the enumerated set of values
\item \lstinline[language=dj]$double$ -- a double-precision floating point number
\end{itemize}

\begin{lstfloat*}
\begin{lstlisting}[language=dj,caption={University database schema definition (Part 2).}, label={lst:uni2}]
::Course     
-> Department
course  : int unsigned   # course number, e.g. 1010
---
course_name :  varchar(200)  # e.g. "Cell Biology"
credits     :  decimal(3,1)  # number of credits earned by completing the course

::Term
term_year : year 
term      : enum('Spring', 'Summer', 'Fall')

::Section 
-> Course
-> Term 
section : char(1)
---
room  :  varchar(12)   # building and room code

::CurrentTerm
---
-> Term

::Enroll
-> Section
-> Student 

::LetterGrade
grade : char(2)
---
points : decimal(3,2)

::Grade 
-> Enroll
---
-> LetterGrade

\end{lstlisting}
\end{lstfloat*}

\begin{figure}
\includegraphics[width=\columnwidth]{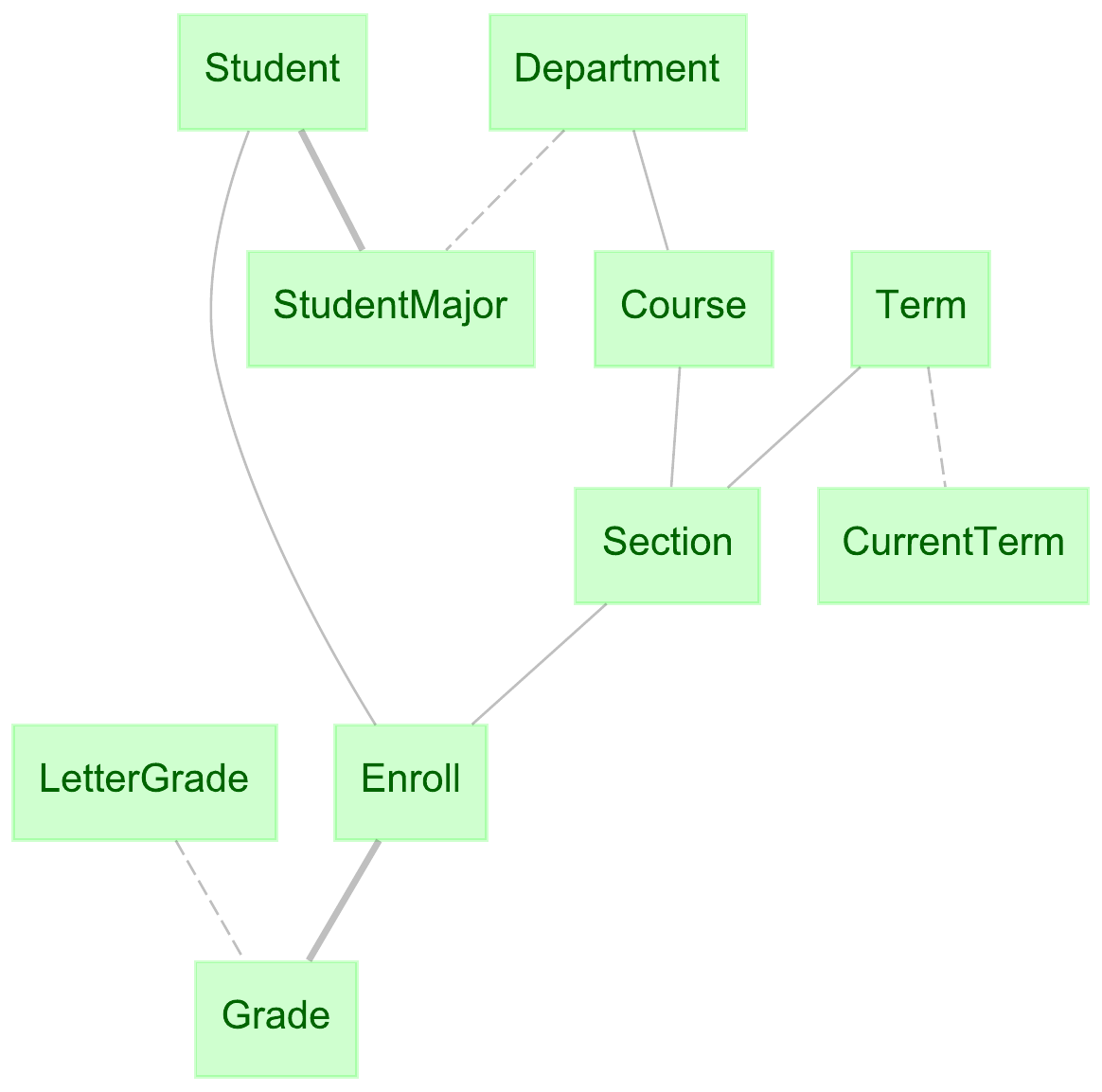}
\caption{The schema diagram of the university database.}
\label{fig:erd}
\end{figure}

\subsection{Primary key}
\subsubsection{Role in entity integrity}\label{sec:entity}
\emph{Entity integrity} is the guarantee made by the data management process that entities from the real world are reliably and uniquely represented in the database system. 
Entity integrity breaks down when the process permits duplicate representations or when a previously created representation cannot be reliably recalled for an external entity.

\emph{Primary keys} are the principal tool for conveying and enforcing entity integrity within the database system. 
Every entity set must have a primary key: a set of attributes whose values, jointly, distinguish elements of the entity set from one another.
No two elements of the same entity set can share the same combination of values of the primary attributes.

In some scenarios, the database system can ensure entity integrity without any measures taken in the external world and the database system can generate primary key values internally and automatically.
One such scenario could be a scientific experiment recording transient entities whose data are entered all at once, never needing to be linked back to the external world again.
Some entities may exist primarily within the database system, \emph{e.g.}\ event logs or application connection sessions; their primary key values may not need to be exposed to the external world either.

For repeated interactions with  persistent entities in the external world, the database system cannot accomplish entity integrity by itself; unique identification of real-world entities must be established by the overall data management process.
This requires measures to assign and maintain a unique and persistent identifying attribute linked to the external entities.
For example, the US Social Security Administration places high value on entity integrity for entities of type ``American Worker'' and goes to great lengths to ensure that, within its overall data management process, the representation of each worker is reliably associated with the same worker and that every worker is associated with the same representation, every time.
The application process for a social security card includes stringent measures to verify one's identity and the use of another person's social security number is prosecuted as a serious offense. 
Many other systems piggyback on the entity integrity effort performed by the Federal Government: social security numbers have gained widespread use as personal identifiers in financial and legal dealings, for example, unrelated to taxes and social security.

In summary, entity integrity is the ability to provide a ready and explicit answer for any entity set to the following question:
\begin{quote}
\em
How does the overall data management process ensure a robust unique mapping between the primary key values of the entity set and the real entities that it claims to represent?
\end{quote}

\subsubsection{Primary key declaration}
In each table declaration (e.g.\ Listings \ref{lst:uni1} and \ref{lst:uni2}), the divider \lstinline$---$ separates the \emph{primary key} attributes above from other attributes below.  
Each entity set must have a primary key although it may comprise multiple attributes. 
For brevity, we refer to the attributes in the primary key as \emph{primary attributes} and all others as \emph{secondary attributes}.

The primary key may comprise a single attribute such as \lstinline$student_id$ in \lstinline$Student$ (Listing \ref{lst:uni1}), multiple attributes such as attributes \lstinline$term_year$ and \lstinline$term$ in entity set \lstinline$Term$, or no attributes at all as in \lstinline$CurrentTerm$ (Listing \ref{lst:uni2}). 
No two elements in \lstinline$Student$ can share the same value of \lstinline$student_id$ and any element can be uniquely identified by \lstinline$student_id$ alone. 
A pair of elements of \lstinline$Term$ can be in the same \lstinline$term_year$ or in the same \lstinline$term$ but cannot be together in both. 

\subsubsection{Phantom attribute $\omega$}\label{sec:phantom}
In explaining the \datajoint data model, we find helpful to introduce the concept of the phantom attribute $\omega$ that is always present in the primary key of all entity sets.  
The domain of $\omega$ comprises one element, say, \lstinline$"1"$, which is also the default value for $\omega$ that is used for inserts.
The phantom attribute is contrived and carries no information but it helps reason through some of the less intuitive aspects of dependencies and query operators.
 
\subsubsection{Singleton entity sets}\label{sec:singleton}
Following the terminology from set theory, the number of elements in an entity set is known as its \emph{cardinality} and the maximum cardinality of an entity set as its \emph{capacity}.
The capacity of an entity set is the cardinality of the domain of its primary key, which is the product of the cardinalities of the domains of all the primary attributes.

A \emph{singleton} entity set is an entity set with the capacity of 1. 
This occurs when the domain of the primary key contains exactly one element.
A particular class of singleton entity sets are those with zero attributes in the primary key, with only $\omega$ remaining.

Singleton entity sets contain singular facts.
For example, \lstinline$CurrentTerm$ in Listing \ref{lst:uni2} is a singleton base entity set whose only element, if present, represents the academic term currently in session.

\subsection{Dependencies}\label{sec:dependencies}
\subsubsection{Referential integrity}\label{sec:referential}
Related data may be distributed across multiple entity sets, creating the potential for dissociations or inconsistencies. 

\emph{Referential integrity} is the guarantee made by the data management process that related data across the database remain present, correctly associated, and mutually consistent.   

Referential integrity is predicated upon entity integrity;
it relies on the use of primary keys for referencing elements and for relating elements between entity sets.

\subsubsection{Declaration of dependencies} 
In \datajoint, referential integrity is supported through the concept of a \emph{dependency}. 
Entity set \lstinline$Dependent$ is said to have a dependency on entity set \lstinline$Ref$ when every element of \lstinline$Dependent$ is defined with respect to a specific element of \lstinline$Ref$. 
A dependency is declared as a separate line of an entity set definition using the right arrow \lstinline$->$ notation:
\begin{lstlisting}[language=dj]
::Dependent
...
-> Ref
...
\end{lstlisting}

An entity set may declare multiple dependencies.  
For example, entity set \lstinline$StudentMajor$ in Listing \ref{lst:uni1} declares two dependencies: one on \lstinline$Student$ from its primary key and another on \lstinline$Department$ as a secondary attribute.

In this context, \lstinline$Dependent$ is said to be the \emph{dependent} entity set and \lstinline$Ref$ the \emph{referenced} entity set; every element of \lstinline$Dependent$ is dependent on its referenced element in \lstinline$Ref$.
An entity set may have dependencies on multiple entity sets.

\subsubsection{Effects of dependencies}\label{sec:effects}
Referential integrity prohibits the creation of a dependent element before creating its referenced element.
Conversely, a referenced element cannot be deleted before all its dependent elements are deleted. 

Declaring a dependency in base entity set \lstinline$Dependent$ on entity set \lstinline$Ref$ performs the following actions:
\begin{enumerate}
\item Add the primary key attributes of \lstinline$Ref$ to the definition of \lstinline$Dependent$ with the same names and datatypes, skipping any that have already been added by other dependencies. Then every element of \lstinline$Dependent$ will contain the primary key value of its referenced element in \lstinline$Ref$, thereby identifying the referenced element.
\item Create the referential constraint that precludes elements in \lstinline$Dependent$ without the corresponding referenced elements in \lstinline$Ref$.
The constraint is enforced by 
\begin{enumerate}
\item prohibiting inserting into \lstinline$Dependent$ of elements unless they reference an existing elements in \lstinline$Ref$,
\item prohibiting deletes of elements from \lstinline$Ref$ that have dependent elements in \lstinline$Dependent$ or propagating such deletes to the dependent elements also,
\item adding caution to updates any attributes values in referenced elements of \lstinline$Ref$, considering that dependent elements in \lstinline$Dependent$ may have been created with consideration for the values of attributes in the referenced elements (See Section \ref{sec:update-caution}).
\end{enumerate}
\item Create an index to accelerate searches in \lstinline$Dependent$ given the primary key values from \lstinline$Ref$ --- if no appropriate index already exists.
\end{enumerate}

The primary key attributes of \lstinline$Ref$ within \lstinline$Dependent$ are known as the \emph{foreign key}.
By construction, foreign key attributes have the same names and datatypes as the primary key of the entity set referenced by the dependency.
Consequently, an entity set can only have one direct dependency on another entity set.  
In other words, the graph of dependencies between entity sets only has single edges between pairs of nodes.  

\subsubsection{Updating referenced elements}\label{sec:update-caution}
\datajoint's stricter form of referential integrity prescribes a cautious procedure for updating attribute values in referenced elements. 
Cautious updates must ensure consistency of dependent elements that are created with consideration of the values of their respective referenced elements.

For example, suppose a student enrolls in a course by adding an element to \lstinline$Enroll$. 
This element  depends on \lstinline$Section$ and \lstinline$Student$ (See Listings \ref{lst:uni1} and \ref{lst:uni2} and Figure \ref{fig:erd}).  
The system must prevent the creation of an element in \lstinline$Enroll$ before creating the corresponding elements in \lstinline$Section$ and \lstinline$Student$. 
Neither should a student or a section be deleted without also deleting corresponding enrollments.
Furthermore, special caution must be taken when updating values of referenced elements.  
If a course section is created with associated enrollments, then updating the number of credits for the course may be restricted.

\subsubsection{Singleton dependencies}
A dependency on a singleton entity set (Section \ref{sec:singleton}) is called a \emph{singleton dependency}. 
Singleton dependencies obey the same rules as other dependencies. 
Since the referenced entity set has no new primary key attributes, a singleton dependency may create a dependency with an empty foreign key.  
Explicit addition of the phantom attribute $\omega$ (Section \ref{sec:phantom}) on both ends of the dependency reduces singleton dependencies to the more conventional use of foreign keys with $\omega$ as the only attribute.

\subsubsection{Derived dependencies}\label{sec:deriv}
In a dependency declaration, the referenced entity set \lstinline$Ref$ is most commonly another base entity set. 
We will refer to such dependencies as \emph{base dependencies}.
All dependencies in the university database in Listings \ref{lst:uni1} and \ref{lst:uni2} are base dependencies.

More generally, \lstinline$Ref$ can be a query expression producing a derived entity set, resulting in a \emph{derived dependency}.
Query expressions are described in Section \ref{sec:query} and we postpone a more detailed discussion of derived dependencies until Section \ref{sec:dep}.

Derived dependencies expand the variety of representable relationships in \datajoint. 
In particular, derived entity sets allow renaming primary key attributes and make it possible to create multiple dependencies between a pair of entity sets, indirectly.

\subsubsection{Primary and secondary dependencies}
Dependencies are categorized as \emph{primary} or \emph{secondary} depending on whether they are declared from the primary or the secondary sections of the base entity set declaration, \emph{i.e.} above or below the divider \lstinline$---$.

For example, \lstinline$StudentMajor$ in Listing \ref{lst:uni1} declares a primary dependency on \lstinline$Student$ and a secondary dependency on \lstinline$Department$.

A primary dependency indicates a closer relationship between the dependent and referenced entity sets than a secondary dependency.
Such a relationship may be described as \emph{defining} or \emph{identifying}. 
The primary dependencies of an entity set become dimensions of the space occupied by its elements. 
Primary dependencies form long-range links across the schema since the primary attributes propagate across multiple steps of dependencies, allowing direct associations between distant entity sets.

Secondary dependencies create a merely \emph{referential relationship}.  
Secondary dependencies become secondary attributes of the dependent entity set with no direct influence on its dependents.

\subsubsection{Distinguishing attributes}\label{sec:distinguish}
The primary dependencies of an entity set define the distinct dimensions of the space occupied by its elements. 
Any primary attributes besides those incorporated through the primary dependencies are called \emph{distinguishing}.  
Each distinguishing attribute defines a new dimension of the space available for the elements of the entity set.  
Distinguishing attributes allow distinguishing multiple elements for each combination of the referenced elements.

Entity sets that have distinguishing attributes are called \emph{distinguished}. 
Distinguished entity set are particularly important because they introduce new semantic dimensions to the schema.
Undistinguished entity sets only relate distinguished entity sets or provide additional information about their elements.

\subsubsection{Homologous attributes}\label{sec:homo}
In a well-designed schema, the same distinguishing attribute will not be defined multiple times in different entity sets.
It will be defined only once and then propagated through dependencies to other entity sets that require it. 
When propagating across a primary dependency, a primary attribute remains primary in the dependent entity set and will thereby continue to propagate to its dependents, often spanning chains of dependencies and creating long-range semantic associations across the schema.

\emph{Homologous attributes} are attributes that trace their origin to the same distinguishing attribute through chains of dependencies.

\emph{Namesake} attributes are attributes in different entity sets that share the same name. 

Using derived dependencies, it is possible to rename attributes in foreign keys so that homologous attributes. 
Therefore, homologous attributes are not always namesake. 
It is then possible to have two or more homologous attributes in the same entity set.

Homologous attributes underpin the logic of query expressions. 

\subsubsection{Relationships}\label{sec:relationship}
In \datajoint, the term \emph{relationship} is used rather generally to describe the effects of particular configurations of dependencies between multiple entity sets. 

For example, the entity set \lstinline$Enroll$ in our University Database  declares primary dependencies on two entity sets \lstinline$Student$ and \lstinline$Section$.
\lstinline$Enroll$  may be thought to denote a many-to-many relationship between \lstinline$Student$ and \lstinline$Section$.

This definition of relationships differs from that in the Entity-Relationship Model where relationships are more narrowly defined as a special type of relations (See Section \ref{sec:erm} and Table \ref{tab:terms}).
For example, in ERM, a student's major might be reasonably modeled as a binary relationship set \lstinline$majors in$ between \lstinline$Student$ and \lstinline$Department$  (Fig.\ \ref{fig:erm-notation}A).
In a \datajoint schema, the same relationship is captured by defining the entity set \lstinline$StudentMajor$ (Fig.\ \ref{fig:erm-notation}B).
The distinction between ERM and \datajoint is mostly semantic but it leads to further notational and conceptual simplifications. 

\subsubsection{Acyclicity}\label{sec:acyclic}
In \datajoint, dependencies between entity sets are not allowed to form cycles; any dependency diagram is an \emph{acyclic directed graph}.
This constraint may seem arbitrary since the model could be straightforwardly extended to allow cyclic dependencies. 
However, we find that a ban on cyclic dependencies greatly simplifies the use of the model with no substantial reduction in its representational power.
With acyclicity, entity sets may be described as ``upstream'' and ``downstream'' of one another according to their topological ordering.
As a consequence, \datajoint databases have a well-defined direction of workflow and are often described as ``data pipelines''.

Cyclic \emph{relationships} can still be captured by an acyclic set of dependencies. 
This commonly requires introducing a downstream entity set to represent the relationship.
For example, the supervisor-subordinate relationship is a classic cyclic relationship because it relates the entity set \lstinline$Employee$ to itself: any employee may be a subordinate to another employee. 
In other relational data models, this could be implemented with a foreign key from \lstinline$Employee$ back into \lstinline$Employee$.
In \datajoint, this cyclic relationship must be modeled by adding a separate downstream entity set \lstinline$Subordinate$ with two dependencies on \lstinline$Employee$ as illustrated in Figure \ref{fig:employee}: a primary dependency to identify the subordinate employee and a secondary dependency to reference his or her manager.

\section{Schema diagrams}\label{sec:diag}
\datajoint's schema diagramming notation takes advantage the conceptual refinements of the data model. 
It is designed to succinctly and clearly communicate all the information necessary to understand the relationships between entities and to construct valid queries.

Figure \ref{fig:erd} illustrates the schema diagramming notation using the university database defined in Listings \ref{lst:uni1} and \ref{lst:uni2}.  

Base entity sets are shown as named rectangular nodes connected with edges representing dependencies.
Primary dependencies are depicted as solid lines and secondary dependencies as dashed lines.
Thick solid lines denote dependencies with the same primary key on both ends, denoting a 1-to-1 relationship: these occur when the dependent entity sets has no distinguishing attributes and no additional primary dependencies.

Schema diagrams are \emph{simple graphs}, as opposed to \emph{multigraphs}, meaning that only one dependency can form between any pair of entity sets.

Because dependencies are acyclic (Section \ref{sec:acyclic}),  schema diagrams are \emph{acyclic directed graphs} with dependencies pointing upward: dependent entity sets are below or \emph{downstream} and referenced entity sets are above or \emph{upstream}.

\section{Data manipulation}\label{sec:manip}
\emph{Data manipulation} refers to operations that change the state of the stored data.  
In \datajoint, the three operations for data manipulation are \emph{insert}, \emph{delete}, and \emph{cautious update}. 

\subsection{Insert}
The \emph{insert} command inserts new elements into an existing base entity set.  
The elements (entities) to be inserted must be fully formed and the entire set is added atomically.
Invalid elements are rejected. 
An element may be invalid if it is incomplete, or if it violates a domain constraint (incorrect value datatype), or if it violates a unique constraint (duplicate value in the primary key), or a referential constraint (no matching entry in a referenced entity set).
When an insert is rejected for any one element, the entire insert set is rejected.

Listing \ref{lst:insert} illustrates the insert command for adding two elements to entity set \lstinline{Student}.
\begin{lstfloat*}
\begin{lstlisting}[language=dj, caption={Inserting two elements into enity set \lstinline{Student}}, label={lst:insert}]
insert Student (student_id, first_name, last_name, sex, date_of_birth, home_address, home_city, home_state, home_zipcode, home_phone):
(1000, 'Rebecca', 'Sanchez', 'F', 1997-09-13, '6604 Gentry Turnpike Suite 513', 'Andreaport', 'MN', '29376', '(250)428-1836'),
(1001, 'Matthew', 'Gonzales', 'M', 1997-05-17, '1432 Jessica Freeway Apt. 545', 'Frazierberg', 'NE', '60485-3810', '(699)755-6306x996')
\end{lstlisting}
\end{lstfloat*}

\subsection{Delete}\label{sec:delete}
The \emph{delete} command removes a subset of elements from a base entity set and the corresponding dependent subsets from any dependent entity sets, cascading recursively down the chains of dependencies.
Deletes cascade recursively to dependent entities according to the foreign keys constraints so that deleting an entity from one entity set triggers the deletion of all matching entities downstream in the data pipeline.
Each invocation of the delete command is executed atomically, no matter how many elements are deleted and how far downstream they cascade along dependency chains.
To specify the subset of elements to remove, delete relies on restriction operator (Sec.\ \ref{sec:restrict}).

The following snippet illustrates the delete command for deleting a subset of elements from \lstinline{Student} and dependent elements in dependent entity sets:
\begin{lstlisting}[language=dj]
delete Student & student_id > 500 & student_id <= 1000
\end{lstlisting}

\subsection{Cautious update}
The \emph{update} operation modifies the values of individual attributes in an entity set.
As explained in Section \ref{sec:referential}, updating values of attributes in place is a precarious proposition from the point of view of referential integrity.
The primary way to change the state of a \datajoint database is through delete and insert operations, which properly enforce referential constraints.
Updates are only performed cautiously in exceptional cases with special consideration of their effects on dependent entity sets. 

The following snippet illustrates the update command for specific attributes of a specific element of \lstinline{Student}:
\begin{lstlisting}[language=dj]
update Student & student_id == 1001:
    home_phone: "(713)555-1040x101",
    home_city: "Houston"
\end{lstlisting}

\section{Queries}\label{sec:query}

\subsection{Expressions and operators}\label{sec:expressions}
In \datajoint, data queries come in the form of \emph{query expressions} that use \emph{query operators} to combine base entity sets into new, \emph{derived entity sets} for precise data queries.
Query expressions may be assigned to variables and used in other expressions but they remain only symbolic representations of the result.
When the state of the database changes, so does the result of the query expression.

\datajoint has five query operators summarized in Table \ref{tab:operators}.
\datajoint's query expressions are \emph{algebraically closed}, meaning that the output of any operator can always be used as input for another. 

\begin{table}[ht]
\rowcolors{1}{gray!20}{white}
\begin{tabu}{|X[1,0.3cm] X[1,1.6cm] X[1,1.6cm]|}
\hline
\rowcolor{HeaderColor}
{\bf operator} & {\bf notation} & {\bf result} \\
{\bf restrict} (Sec.\ \ref{sec:restrict})  & \makecell{\lstinline$A \& cond$\\ \lstinline$A \\ cond$}   & The subset of all elements of \lstinline$A$ that meet (or do not meet) condition \lstinline$cond$.\\
{\bf join} (Sec.\ \ref{sec:join}) & \lstinline$A * B$ & The combination of matching elements of \lstinline$A$ and \lstinline$B$. \\
{\bf project} (Sec.\ \ref{sec:proj}) & \lstinline$A.proj(...)$ & Select, rename, and calculate attributes for \lstinline$A$. \\
{\bf aggregate} (Sec.\ \ref{sec:aggr}) & \lstinline$A.aggr(B, ...)$ & Calculate new attributes for \lstinline$A$ using aggregation operations on attributes of \lstinline$B$. \\
{\bf union} (Sec.\ \ref{sec:union}) & \lstinline$A + B$ & The set of elements that are in either \lstinline$A$ or \lstinline$B$ or both. \\ 
\hline
\end{tabu}
\caption{\datajoint query operators.}
\label{tab:operators}
\end{table}

\subsection{Operational entity normalization}\label{sec:operational}
\datajoint extends the principle of \emph{entity normalization} (Section \ref{sec:norm}) to query expressions.
Every derived entity set must meet all the criteria of a valid entity set. 
Given well-defined entity set as inputs, any \datajoint operator must produce a well-defined entity set as output with a readily identifiable entity type and primary key.

We refer to this property of \datajoint expressions as \emph{operational entity normalization}; 
it sets \datajoint apart from other relational query languages, which limit the formal application of concepts of normalization and entity integrity to base entity sets.

\subsection{Uniform logic of binary operators}\label{sec:match}
Binary query operators are those that use two entity sets as inputs.
They are \emph{restrict} (when restricting by an entity set), \emph{join}, \emph{aggregate}, and \emph{union}.
Binary operators work by matching values of attributes across their two operands.
The choice of the attributes to match defines the semantics of the operation. 

Other relational query languages provide multiple ways to specify the choice of matching attributes.
For example, the choice may be explicit, requiring a careful examination of the attribute semantics and resulting in unwieldy queries.
Another choice are \emph{natural joins}, wherein namesake attributes are matched on the assumption that they are semantically related.
Natural joins lead to frequent mistakes when unrelated attributes happen to share the same names.

\datajoint offers a single uniform method for defining the semantics of binary operators through the notion of \emph{homologous namesake attributes} as defined in Section \ref{sec:homo}.
This semantic refinement avoids many of the pitfalls of natural joins and applies more generally to all binary operators.  
All binary query operators work by restricting the result to those element pairs whose homologous namesake attributes assume equal values.

For example, the entity sets \lstinline$StudentMajor$ and \lstinline$Enroll$ (Listings \ref{lst:uni1} and \ref{lst:uni2} Figure \ref{fig:erd}) have homologous namesake attributes \lstinline$student_id$ and \lstinline$dept$, which they acquire through a series of dependencies from the primary keys of \lstinline$Student$ and \lstinline$Department$, respectively.  

Then, in the course of the binary operators 
\begin{description}
\item[join]  \lstinline$StudentMajor * Enroll$ 
\item[restrict] \lstinline$StudentMajor & Enroll$ and \\ \lstinline$StudentMajor \ Enroll$ 
\item[aggregate] \lstinline$StudentMajor.aggr($\\
   \lstinline$Enroll, n: count())$ 
\end{description}
 elements of \lstinline$StudentMajor$ are paired with elements of  \lstinline$Enrolled$ where values of  \lstinline$student_id$ and \lstinline$dept$ are equal.

Universal sets allow additional control of attribute matching in binary operators and are described in detail in Section \ref{sec:u}.

\subsection{Restrict}\label{sec:restrict}
The \emph{restriction} expression \lstinline$A & cond$ produces an entity set comprising the subset of entity set \lstinline$A$ whose elements  meet the condition \lstinline$cond$. 
The output is an entity set with the same primary key, the same entity type, and the same attributes as \lstinline$A$.
The \emph{exclusion} operator \lstinline{\} yields the complementary subset.
Restrict and exclude have similar semantics; we use the term \emph{restriction} to refer to both operators, considering exclude as only a variant of restrict.

In relational algebra, restriction is also known as \emph{selection}.  
We use the term \emph{restriction} to avoid confusion with SQL's \lstinline$SELECT$ statement, which does much more than restrict and where only the \lstinline$WHERE$ clause performs restriction.

As described in section \ref{sec:delete}, the delete command uses restrictions to specify the subset to be deleted.

Variants of restriction depend on the form of \lstinline$cond$ as described below.

\subsubsection{Restriction by attribute conditions}
One form of restriction is when \lstinline$cond$ is an explicit condition on attribute values.  
Such conditions may include arithmetic operations, functions, range tests, etc.
Some examples are listed in Listing \ref{lst:res1}.
\begin{lstlisting}[language=Python, caption={Restrictions by attribute conditions.}, label={lst:res1}]
# Students from Texas
Student & home_state == "TX"
# Students not from Texas
Student & home_state <> "TX"
Student \ home_state == "TX"
# Male students not from Texas
Student & sex == "M" \ home_state == "TX"
\end{lstlisting}

As with all query expressions, the result of restriction can be assigned to a variable and used in subsequent queries (e.g. Listing \ref{lst:res2}).
\begin{lstlisting}[language=Python, caption={Assignment and use of relational variables.}, label={lst:res2}]
Millennial = Student & 
   date_of_birth >= "1980-01-01" & 
   date_of_birth < "2001-01-01"

MillennialMale = Millennial & sex == "M"
\end{lstlisting}

\subsubsection{Restriction by a key-value mapping}
Restriction may be done by a \emph{key-value mapping} in the form \lstinline${key1: value1, ..., keyN: valueN}$.
The key-value pairs where the key is an attribute in \lstinline$A$ are treated as equality conditions. 
All other key-value pairs are ignored.

For example, the two query expressions in Listing \ref{lst:res-map} are equivalent given the current definition of \lstinline$Student$.
\begin{lstlisting}[language=Python, caption={Equivalent expressions using restrictions by a mapping and by attribute conditions.  The condition on \lstinline$dept$ is ignored because it is not an attribute in \lstinline$Student$.}, label={lst:res-map}]
# restriction by a key-value mapping
Student & {
   first_name: "Alice", 
   last_name: "Cooper", 
   dept: "MATH"}

# equivalent attribute condition 
Student & first_name == "Alice" & 
    last_name == "Cooper"
\end{lstlisting}

Accordingly, a restriction by a mapping yields the original unrestricted set when when none of the mapping's keys match any attributes in the entity set or when the mapping has no keys: \lstinline$A & {}$.
Conversely, exclusion by such a mapping yields the empty set.  
This is illustrated in Listing \ref{lst:res-empty-map}.

\begin{lstlisting}[language=Python, caption={Restriction by a mapping with no matching keys}, label={lst:res-empty-map}]
# No restriction
Student & {}
Student & {dept: "MATH"}

# Empty set
Student \ {}
Student \ {dept: "MATH}
\end{lstlisting}

\subsubsection{Restriction by a collection}
When \lstinline$cond$ is a collection (\emph{e.g.}\ a list or a set) of conditions, then the conditions are applied by logical disjunction (logical {OR}):
The result of \lstinline$A & [cond1, ..., condN]$ contains all elements of \lstinline$A$ that meet \emph{any} of the conditions \lstinline$cond1, ..., condN$.

\begin{lstlisting}[language=Python, caption={Restrictions by a collection of conditions.}, label={lst:res-list}]
# Students from Oklahoma, New Mexico, or Texas
Student & [
   home_state == "OK", 
   home_state == "NM", 
   home_state == "TX"] 

# An equivalent restriction using the IN operator
Student & home_state in ["OK", "NM", "TX"]
\end{lstlisting}

Accordingly, \lstinline$A & []$ is empty whereas \lstinline$A \ []$ yields the original set \lstinline$A$.

\subsubsection{Restriction by an And-collection}
The special function \lstinline$And$ represents logical conjunction (logical {AND}) so that \lstinline$A & And([cond1, ..., condN])$ is equivalent to \lstinline$A & cond1 & ... & condN$.

Accordingly, the result of \lstinline$A \ And([])$ is empty and \lstinline$A & And([])$ yields the original set \lstinline$A$.

\subsubsection{Restriction by a negation}
The special function \lstinline$Not$ represents logical negation so that \lstinline$A & Not(cond)$ is equivalent to \lstinline$A \ cond$.

De Morgan's laws can be applied to derive logically equivalent expressions.  
For example, \lstinline$A \ [a, b]$ is equivalent to \lstinline$A \ a \ b$ whereas \lstinline$A \ And([a, b])$ is equivalent to \lstinline$A & [Not(a), Not(b)]$ and so forth.

\subsubsection{Restriction by an entity set}
When \lstinline$cond$ is another entity set, the result of \lstinline$A & cond$ comprises all elements from \lstinline$A$ for which there exists an element with equal values of the homologous namesake attributes in \lstinline$cond$.  
All other attributes are ignored.  

Since homologous attributes are semantically related, restriction by an entity set yields intuitive and well-defined results.

Listing \ref{lst:res-set} illustrates some queries with restrictions by entity sets using our university database.
\begin{lstlisting}[language=Python, caption={Queries with restrictions by entity sets.}, label={lst:res-set}]
# Students who have taken classes
Student & Enroll
# Students who have not taken classes
Student \ Enroll
# Students who have not selected a major
Student \ StudentMajor
\end{lstlisting}

\begin{lstlisting}[language=Python, caption={Composite restrictions.}, label={lst:res-comp}]
# Students who have taken Biology classes but no MATH courses
Student & 
  (Enroll & dept == "BIOL") \ 
  (Enroll & dept == "MATH")

# Students who are not taking courses in the current term
Student \ (Enroll & CurrentTerm)

# Students who have taken classes and have chosen a major
Student & Enroll & StudentMajor

# Students who have taken classes in their major
Student & (Enroll & StudentMajor)

# Students who are taking classes outside their major in the current term
Student & (Enroll \ StudentMajor & CurrentTerm)

#Students who have taken classes or have chosen a major
Student & [Enroll, StudentMajor]
\end{lstlisting}

In relational algebra, the closest equivalent to \datajoint's restriction and exclusion operators by an entity set are called, respectively, \emph{semijoin} and \emph{antijoin}.
We deprecate these confusing terms because these operators are restriction-like and not join-like in their semantics.

\subsection{Join}\label{sec:join}
The result of the join operator \lstinline$A * B$ combines \lstinline$A$ and \lstinline$B$ into a single entity set.
The primary key of a join is the union of the primary key attributes of its operands. 
The entity type of the result is the pairing of the entity types of the operands.  

\datajoint considers entity sets \lstinline$A$ and \lstinline$B$ \emph{joinable} when all their namesake attributes are also homologous.
An attempt to join unjoinable entity sets will raise an exception.
Any two entity sets can be made joinable by applying appropriate projection operators (Section \ref{sec:proj}) to its operands to remove or rename namesake attributes that should not be involved in the logic of the join.
In addition to stricter semantic clarity, join compatibility ensures unambiguous unique attribute names in the result.

For example, Listing \ref{lst:join1} illustrates queries that combine data from multiple base entity sets into a single derived entity set.
\begin{lstlisting}[language=Python, caption={Combining entities.}, label={lst:join1}]
# Grade point values
Grade * LetterGrade

# Graded enrollments with complete course and student information
Student * Enroll * Course * Section * Grade * LetterGrade
\end{lstlisting}

Listing \ref{lst:join2} illustrates complex queries that include joins.
\begin{lstlisting}[language=Python, caption={Join in expressions.}, label={lst:join2}]
# Students with ungraded courses in current term
Student & (Enroll * CurrentTerm \ Grade)

# Enrollments before students' date of birth
Student * Enroll & (term_year <= date_of_birth)
\end{lstlisting}

\subsection{Projection}\label{sec:proj}
The projection operator \lstinline$A.proj(attr1, ..., attrN$ is a unary operator on the input entity set \lstinline$A$.  
Its output is an entity set with the same entity type and the same number of elements as \lstinline$A$ with some attributes renamed, some secondary attributes omitted, or new calculated attributes introduced.
A single projection operator can perform any combination of all three of these functions in a single invocation. 

\subsubsection{Attribute selection}
Projection omits all secondary attributes except those listed as arguments;
it retains the primary attributes to guarantee that the output is a well-defined entity set.

Listing \ref{lst:select} shows examples of attribute selection.
\begin{lstlisting}[language=Python, caption={Selecting  attributes.}, label={lst:select}]
# Student id, firs name, and last name
Student.proj(first_name, last_name)

# Student id only
Student.proj()
\end{lstlisting}

\subsubsection{Renamed attributes}
When the attribute specification in a projection call is the form \lstinline$new: old$, this renames an existing attribute in the relation from \emph{old} to \emph{new}.

For example, the result of the query in Listing \ref{lst:ren1} renames attribute \lstinline$dept$ into \lstinline$major$.
\begin{lstlisting}[language=Python, caption={Renaming attributes.}, label={lst:ren1}]
# rename dept into major
StudentMajor.proj(major: dept)
\end{lstlisting}

Renaming is commonly  used to control the semantics of binary operators such as joins, restrictions, and aggregations.

For example queries in Listing \ref{lst:rename} differ in meaning thanks to the renaming of the matching attribute \lstinline$dept$.
In queries 1 and 3, \lstinline$dept$ is among the matching attributes. 
In \lstinline$Enroll$ and \lstinline$Grade$, \lstinline$dept$ refers to the department offering the course; in \lstinline$StudentMajor$, it denotes the students' major departments. 

In queries 2, 4, and 5, \lstinline$dept$ is renamed into \lstinline$major$; it does not restrict the enrollment records and \lstinline$major$ becomes a distrinct attribute in the output of the joins. 
In query 5, the two attributes \lstinline$major$ and \lstinline$dept$ are used in a further restriction of the join result.

\begin{lstlisting}[language=Python, caption={Renaming attributes.}, label={lst:rename}]
# 1. Enrollments by students in the same 
# major as the department offering course 
Enroll & StudentMajor

# 2. Enrollments by students who have 
# chosen a major
Enroll & StudentMajor.proj(major: dept)

# 3. Grades in courses within student's major 
Grade * StudentMajor

# 4. Grade and major information 
Grade * StudentMajor.proj(major: dept)

# 5. Grade outside chosen major
Grade * StudentMajor.proj(major: dept) & major != dept
\end{lstlisting}

Renaming is often necessary when joining one entity with itself to obtain pairings of elements from the same entity set.
For example, the query in Listing \ref{lst:bday} pairs students but it must rename the primary key attribute in one of the operands to allow pairing and it must rename or project out secondary attributes. 
\begin{lstlisting}[language=Python, caption={Joining an entity set with itself}, label={lst:bday}]
# pairs of students with the same birthdays
Student * Student.proj(
   student_id2: student_id, 
   date_of_birth2: date_of_birth) & 
      student_id < student_id2 & 
      date_of_birth = date_of_birth2
\end{lstlisting}

Renaming may also be used to make joinable entity sets that are non-joinable due to the presence of non-homologous namesake attributes.
By design, our university database lacks unjoinable entity sets but imagine if both \lstinline$Department$ and \lstinline$Enroll$ had secondary attributes \lstinline$room$ that were non-homologous, \emph{i.e.}\ not referencing the same original definition.  
That would make these two entity sets non-joinable because the attribute \lstinline$room$ would be ambiguous in the output.
To make them joinable again, the offending attribute would need to be renamed before the join as in 
\begin{lstlisting}
Enroll.proj(classroom: room, ...) * Department
\end{lstlisting}
Here the ellipsis \lstinline$...$ instructs to include all the remaining attributes. 

\subsubsection{Calculated attributes}
Projection can also be used to calculate new attributes for each element of the entity set based on the existing attributes of the same element.
The new attribute name and the calculations are specified as  \lstinline$attr: expression$.
Calculated attributes are always secondary.
 
For example, the first query in Listing \ref{lst:calc}, the attribute \lstinline$total$ calculates the total points earned by students in each course as the product of attributes \lstinline$credits$ and \lstinline$points$ that come from \lstinline$Course$ and \lstinline$LetterGrade$, respectively, for each entry in \lstinline$Grade$.
As an example of a more precise query, the second query adds  (\lstinline$course_name$) to the output and restricts to the current term.
\begin{lstlisting}[language=Python, caption={Extension: calculated attributes.}, label={lst:calc}]
# Total grade points
(Grade * Course * LetterGrade).proj(
    total: points * credits)

# Total grade points in current term
(Course * Grade * LetterGrade).proj(
    course_name, 
    total: points * credits) & CurrentTerm
\end{lstlisting}

\subsection{Aggregation}\label{sec:aggr}
Aggregation is a generalization of the projection operator. 
Operator \lstinline$A.aggr(B, ...)$ can do anything that operator \lstinline$A.proj(...)$ can.  
In fact, without the argument \lstinline$B$, the two operators are exactly equivalent.
Aggregation allows adding calculated attributes to each element of \lstinline$A$ based on aggregation functions over attributes in the matching  subsets of elements of \lstinline$B$.

Aggregation functions include \lstinline$count$, \lstinline$sum$, \lstinline$min$, \lstinline$max$, \lstinline$avg$, \lstinline$median$, \lstinline$percentile$, \lstinline$stddev$, \lstinline$var$, and others. 
Aggregation functions cannot be used anywhere else other than in definitions of new attributes in the \lstinline$aggr$ operator.

Just like in projection, the output of aggregation \lstinline$A.aggr(B, ...)$ has the same entity class, the same primary key, and the same number of elements as \lstinline$A$.
Just like in projection, the primary key attributes are always included in the output but may be renamed.

From the point of view of SQL and relational algebra, aggregation is equivalent to the left outer join of \lstinline$A$ and \lstinline$B$ on their homologous namesake attributes followed by the \lstinline$GROUP BY$ operation on the primary key attributes of \lstinline$A$.

Generally, \lstinline$A$ and \lstinline$B$ may have namesake attributes that are not homologous but then they may not be included in the output due to the ambiguity.  
Making \lstinline$A$ and \lstinline$B$ joinable avoids any ambiguity in referencing attributes.

For example, Listing \ref{lst:aggr1} illustrates aggregation queries for the university database.
\begin{lstlisting}[language=Python, caption={Calculate summary statistics.}, label={lst:aggr1}]
# Number of students in each course section
Section.aggr(Enroll, n: count())

# Average grade in each course
Course.aggr(Grade * LetterGrade, 
    avg_grade: avg(points))
\end{lstlisting}

In Listing \ref{lst:aggr2}, queries make use of variables storing derived entity sets and then used in new queries.
\begin{lstlisting}[language=Python, caption={Calculate summary statistics using variables to store subqueries.}, label={lst:aggr2}]
# 1. Number of students enrolled per section
enrolled = Section.aggr(Enroll, n: count())

# 2. Number of students graded per section
graded = Section.aggr(Grade, m: count())

# 3. Fraction of enrolled students with grades
fraction = (enrolled * graded).proj(
    frac: m/n)

# 4. Student GPA
grades = Course * Grade * LetterGrade
GPA = Student.aggr(grades,
    gpa: sum(points * credits)/
         sum(credits))

# 5. Average GPA across university 
# by student major
Department.aggr(StudentMajor * GPA, 
    avg_gpa: avg(gpa))

# 6. GPA in current term
Student.aggr(grades & CurrentTerm, 
    gpa: sum(points * credits) / sum(credits))
\end{lstlisting}

\subsection{Union}\label{sec:union}
The output of the union operator \lstinline$A + B$ contains all elements of both \lstinline$A$ and \lstinline$B$.  
From the perspective of entity normalization, union requires that its operands \lstinline$A$ and \lstinline$B$ belong to the same entity type with the same primary key with homologous attributes.
In the simplest case, without any secondary attributes, the result is the simple set union.

In the presence of secondary attributes, the secondary attributes must have the same names and datatypes in the two operands. 
Additionally, the two operands must be disjoint, without any duplicate values of the primary key across both inputs.
These definitions avoid any ambiguity of attribute values or entity identity. 

\subsection{Universal sets}\label{sec:u}
\datajoint allows using \emph{universal sets} to define virtual entity sets with arbitrary primary key structure for use in query expressions. 
A universal set denotes the set of all possible elements with given attributes of any possible datatypes.
Let \lstinline$U(attr1, ..., attrN)$ denote the entity set with attributes \lstinline$attr1$, $\ldots$, \lstinline$attrnN$.
In the \datajoint query language, \lstinline$U$ is a reserved word for defining universal sets.
All attributes of a universal set are considered primary.
When the attribute list is empty, \lstinline$U()$ has exactly one element with zero attributes; or it may be thought to have the single phantom attribute $\omega$ as described in Section \ref{sec:phantom}.

Universal sets become useful for defining entity sets as necessary for query expressions when no suitable base entity sets exist.
In expressions, their attributes are allowed to be matched to any namesake attributes without the requirement of being homologous.

Imagine we needed to query the university database for the complete list of students' home cities and get the number of students from each city.
The university database lacks the entity set for cities and states. 
The virtual entity set \lstinline$U(home_city, home_state)$ can play the role of the base entity set for cities as illustrated in Listing \ref{lst:city}.
Although the virtual entity set is nameless, its primary key implies the class type it represents.

\begin{lstlisting}[language=Python, morekeywords={avg, U}, caption={Using universal set to create virtual entity sets for City and State.}, label={lst:city}]
# All home cities of students 
U(home_city, home_state) & Student 

# Total number of students from each city
U(home_city, home_state).aggr(
            Student, n: count())

# Total number of students from each state
U(home_state).aggr(Student, n: count())

# Total number of students in the database
U().aggr(Student, n: count())
\end{lstlisting}

Note that the result of aggregation on a universal set is restricted to the elements with matches in the aggregated set. 
In other words, \lstinline$X.aggr(A, ...)$ is interpreted as \lstinline$(X & A).aggr(A, ...)$ when \lstinline$X$ is a universal set.

Universal sets should be used sparingly when no suitable base entity sets already exist.
Resorting to a universal set effectively states, ``I am defining a new entity set for this query because no other appropriate entity set exists.''
In some cases, it may be preferable to define new base entity sets to make queries clearer and more semantically constrained.

For example, we could define base entities \lstinline$City$ and \lstinline$State$ and then use them in \lstinline$Student$ as follows.
\begin{lstlisting}[language=dj, label={lst:state}, caption={Adding entity sets for \lstinline$State$ and \lstinline$City$ to avoid using universal sets in queries.}]
# singleton entity set for the United States 
::USA 

::State
-> USA
state char(2) :  two-letter acronym 
---
state_name : varchar(20) #  full state name

::City
city : varchar(30)
-> State

::Student     
student_id : int unsigned   # university ID 
---
first_name      : varchar(40)
last_name       : varchar(40)
sex             : enum('F', 'M', 'U')
date_of_birth   : date
home_address    : varchar(200) # street address
-> City 
\end{lstlisting}

Then the queries in Listing \ref{lst:city} could be replaced with equivalent queries in Listing \ref{lst:city2}.
The explicit definition of \lstinline$City$ and \lstinline$State$ make the queries clearer but requires creating and populating extra base entity sets, which are avoided by using universal sets.
\begin{lstlisting}[language=Python, morekeywords={avg, U}, caption={Queries equivalent to Listing \ref{lst:city} made possible by adding more base entity sets in Listing \ref{lst:state}.}, label={lst:city2}]
# All home cities of students 
City & Student 

# Total number of students from each city
City.aggr(Student, n: count())

# Total number of students from each state
State.aggr(Student, n: count())

# Total number of students in the database
USA.aggr(Student, n: count())
\end{lstlisting}

Listing \ref{lst:u2} provides a few more query examples where, in the absence of a suitable base entity, a universal entity is used.

\begin{lstlisting}[language=Python, morekeywords={avg, U}, caption={Aggregation by universal entities.}, label={lst:u2}]
# Total number of unique last names
U().aggr(U(last_name) & Student, n: count())

# Number of students from each state by sex
U(home_state, sex).aggr(Student, 
    n: count())

# Student GPA, retaining the sex attribute
GPA = Student.aggr(sex,
    Course * Grade * LetterGrade,
    gpa: sum(points * credits) /
         sum(credits))

# Average GPA by sex
GpaBySex = U(sex).aggr(GPA, 
    avg_gpa: avg(gpa))
\end{lstlisting}

\subsection{No outer joins}
\datajoint does not support outer joins as they contradict the principle of operational entity normalization.

In relational algebra and SQL, an \emph{outer join} of relations \lstinline$A$ and \lstinline$B$ adds to its output all the elements of \lstinline$A$ (\emph{left outer join}) or  \lstinline$B$ (\emph{right outer join}), or both (\emph{full outer join}) in addition to the entire result of the inner joins. 
Outer joins contradict the concept of entity normalization (Sections \ref{sec:norm} and \ref{sec:operational}) because it mixes entity types with different primary keys into a single relation.
The primary key of an outer join is poorly defined and may have nulls in it, breaking entity integrity.
Outer joins are commonly used to prepare input into a \lstinline$GROUP BY$ operation.
\datajoint's \lstinline$aggr$ operator accomplishes both steps (left join and \lstinline$GROUP BY$) as a single operator, always observing entity normalization in its operations and reducing the need for outer joins.

\section{Advanced schema definition}\label{sec:def2}

\subsection{Uses of derived dependencies}\label{sec:dep}
Section \ref{sec:deriv} introduced the concept of derived dependencies.  
In a derived dependency \lstinline$-> Ref$, the referenced entity set is a query expression rather than a base entity set.
Derived dependencies enrich the representational power of \datajoint schemas.

Importantly, the query expression in the derived dependency may not rely on downstream entity set to prevent cyclic dependencies.

Each dependency \lstinline$-> Ref$ references only the primary key of \lstinline$Ref$, making the dependency equivalent to the empty projection \lstinline$-> Ref.proj()$.

\subsubsection{Dependency on a restriction} 
Dependency on a restriction \lstinline$-> Ref & cond$ has the effect of preventing inserts of elements unless they reference an element that meets the condition \lstinline$cond$.   
The restriction has no effect on deletes or on query expressions involving entity sets from both side of the dependency.
Dependencies on restricted entity sets differ little from dependencies on the unrestricted entity set. 
In the schema diagram, the restricted entity entity set is not shown separately and does not have any special notation.

\subsubsection{Dependency on a join}
Conceptually, a dependency on the join \lstinline$-> A * B$ is exactly equivalent to the two restricted dependencies
\begin{lstlisting}
-> A & B
-> B & A
\end{lstlisting}

In the case when \lstinline$A$ and \lstinline$B$ share no secondary homologous namesake attributes, the dependency is exactly equivalent to the unrestricted separate dependencies
\begin{lstlisting}
-> A 
-> B
\end{lstlisting}

Due to these equivalences, schema diagrams do not require special notation for dependencies on joins, showing them as separate dependencies. 

\subsubsection{Dependency on a projection}
Dependencies only use the primary key of the referenced entity set. 
Hence, the only relevant application of projection is for the renaming of primary key attributes --- which is also one of the most useful applications of derived dependencies.

\begin{figure}
\centering
\includegraphics[scale=0.8]{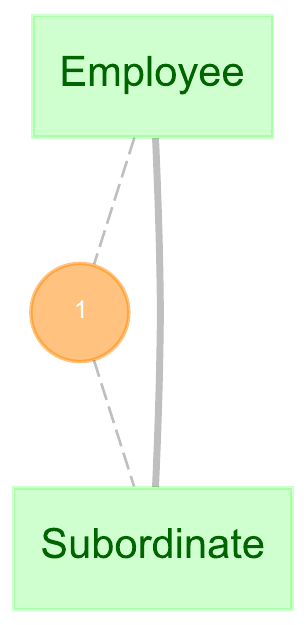}
\caption{
The schema diagram for the manager-subordinate relationship as defined in Listing \ref{lst:employee}.
}
\label{fig:employee}
\end{figure}

For example, imagine we needed to represent the supervisor-subordinate relationship discussed in Section \ref{sec:relationship}.  
Suppose the entity set \lstinline$Employee$ is already defined with \lstinline$emp_id$ as the primary key. 
Then the entity set \lstinline$Subordinate$ could define two dependencies on \lstinline$Employee$ as defined in Listing \ref{lst:employee} and Figure \ref{fig:employee}:
\begin{lstlisting}[language=dj,label=lst:employee,caption={Entity set \lstinline$Subordinate$ expressing the manager-subordinate relationship using a derived dependency.}]
::Subordinate
-> Employee
---
-> Employee.proj(manager: emp_id)
\end{lstlisting}

The resulting entity set will have two attributes: the primary key \lstinline$emp_id$ for the employee herself and the secondary attribute \lstinline$manager$.  Both attributes are foreign key attributes referring \lstinline$Employee$.

As another example, suppose we need to represent prerequisite courses for each course in our university database.  
Listing \ref{lst:prereq} defines entity set \lstinline$Prerequisite$ to represent the prerequisite relationship between courses.
\begin{lstlisting}[language=dj,label=lst:prereq, caption={Entity set \lstinline$Prerequisite$}.]
::Prerequisite
-> Course
-> Course.proj(
    pre_dept: dept, pre_course: course)
\end{lstlisting}

\begin{figure}
\centering
\includegraphics[scale=0.8]{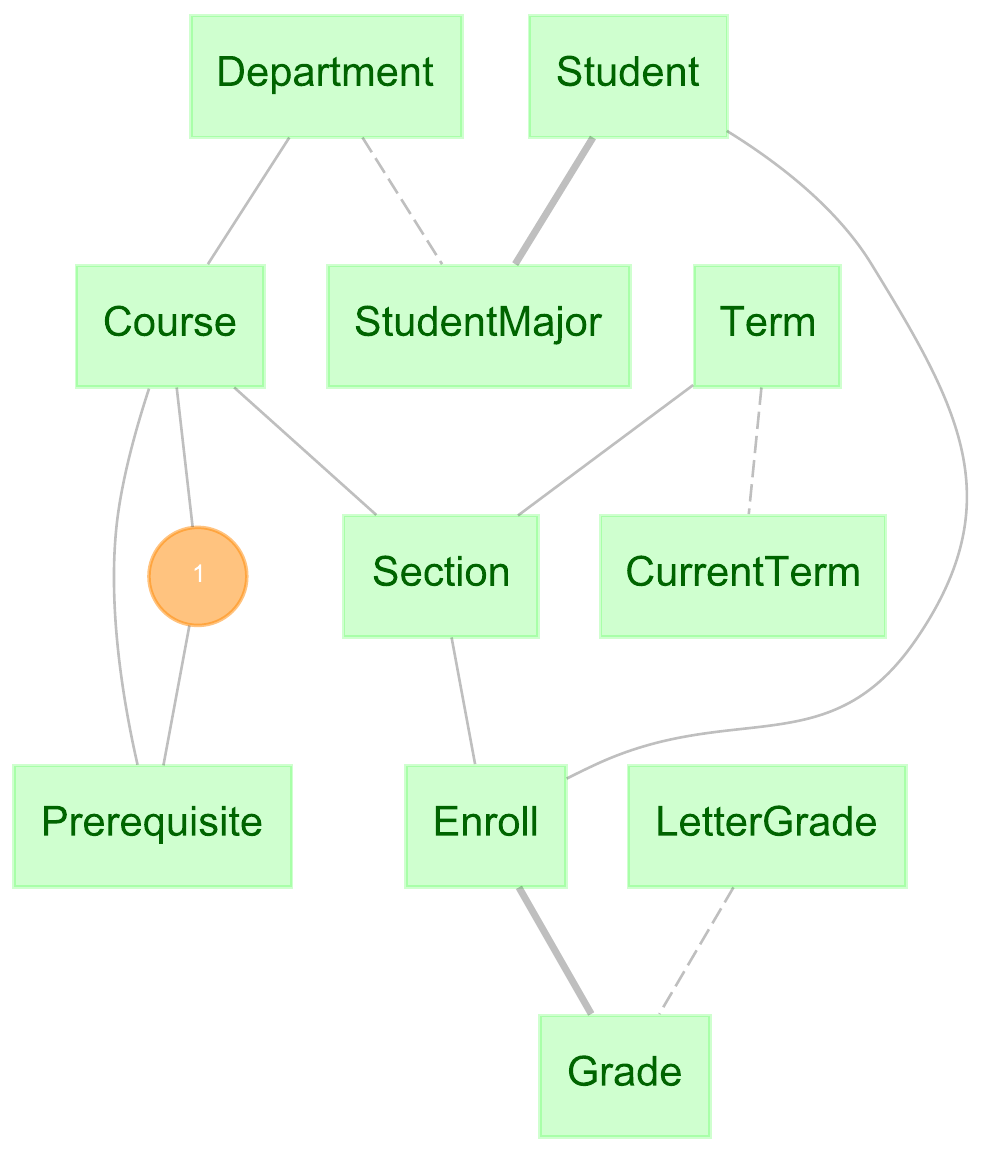}
\caption{
The schema diagram of the university schema including entity set \lstinline$Prerequisite$ defined in Listing \ref{lst:prereq}.
The derived entity set for the derived dependency is shown as a separate node in orange.
}
\label{fig:prereq}
\end{figure}

In the schema diagram, projection is shown as a separate node (Figure \ref{fig:prereq}).

\subsubsection{Dependency on a union}
Dependency on a union finds many uses when the dependent entity set depends on entities in \emph{either} of two entity sets. 
For example, imagine that the university database also contains university employees and library members must be either students \emph{or} employees.
This might be implemented with the base entity sets defined in Listing \ref{lst:uni-3}, which provides two semantically equivalent definitions of \lstinline$LibraryCard$: one with a union dependency and the other with a restricted dependency.

\begin{lstlisting}[language=dj,caption={Alternative university database schema definition with employees and library cards.},label={lst:uni-3}]
::Person
person_id : int unsigned   # university ID 
---
first_name      : varchar(40)
last_name       : varchar(40)
sex             : enum('F', 'M', 'U')
date_of_birth   : date
...

::Student
-> Person
---
admission_date  : date

::Employee
-> Person
---
hire_date : date

::LibraryCard
-> Student.proj() + Employee.proj()
---
expiration  : date 

# Alternative nearly equivalent definition
::LibraryCard
-> Person & [Student, Employee]
---
expiration  : date 
\end{lstlisting} 

\subsection{Queries with derived dependencies}
Explicit use of query expressions to define dependencies simplifies the formulation of valid queries. 
The query expressions defining the dependency directly translate into queries or subqueries relating the dependent and referenced entity sets.

Listing \ref{lst:preq} illustrates how the query expression used in the definition of the dependency in \lstinline$Prerequisite$  is reused to form a data query.
\begin{lstlisting}[language=dj,caption={Reusing the query expressions from the dependency declaration in Listing \ref{lst:prereq} in a data query.}, label={lst:preq}]
# Course BIOL-1010
c = Prerequisite & {dept: "BIOL", course: 1010})

# Prerequsite details for the course
Course.proj(
   pre_dept: dept, pre_course: course) & c

# Sections of prerequisites in current term
(Section & CurrentTerm).proj(
   pre_dept: dept, pre_course: course) & c
\end{lstlisting}

\subsection{Dependency properties}
Dependencies may specify additional properties \emph{unique} and \emph{nullable}. 
They are specified in square brackets in front of the query expression.
\begin{lstlisting}[language=dj]
-> [unique, nullable] Ref
\end{lstlisting}
By default, the dependency is not nullable and is not unique.

\subsubsection{Unique dependencies}
With a unique dependency, each element of the referenced entity set cannot be referenced by more than one element in the dependent entity set. 
Effectively, this defines a uniqueness constraint on the foreign key attributes and enforces a 1:1 relationship between the entity sets even when the dependency is not primary.

\subsubsection{Nullable dependencies}
Only secondary dependencies may be declared with the \emph{nullable} option.  
This indicates an optional dependency where the dependent foreign key attributes may be omitted, \emph{i.e.}\ set to \lstinline{null}.
A unique dependency may also be nullable, in which case the null references are exempted from the  uniqueness constraint.

\subsection{Master-part relationship}\label{sec:master}
Let \emph{compositional integrity} denote the guarantee made by the data management process that entities composed of multiple parts always appear in their complete form.

In \datajoint, compositional integrity is formalized through the notion of a \emph{master-part relationship}. 
In a master-part relationship, one base entity set is designated as the \emph{master}.
A master may have one or several dependent \emph{part} entity sets. 
In \datajoint, the names of part entity sets are prefixed with the name of their master entity set. 

For example, Listing \ref{lst:master} defines a database schema with the master-part relationship between \lstinline$Order$  and \lstinline$Order.Item$.  
\begin{lstlisting}[language=dj,label={lst:master},caption={A schema containing a master-part relationship between \lstinline{Order} and \lstinline{Order.Item}}.]
::Customer
customer : char(16)  # customer id
---
customer_name : varchar(128)
customer_address : varchar(255)

::Product
product : char(16)  # SKU number
--- 
product_name : varchar(120)
product_description : varchar(4000)

::Order
order  : int
---
-> Customer
order_datetime : datetime 
    
::Order.Item
-> master
item : int
---
-> Product
unit  :  char(10)  # unit of measurement
price : decimal(7,2)  # price per unit
quantity : decimal(7,3)  

::Shipment
-> Order
---
tracking_number  : char(36)
ship_datetime : datetime
\end{lstlisting}

\begin{figure}
\centering
\includegraphics[scale=0.8]{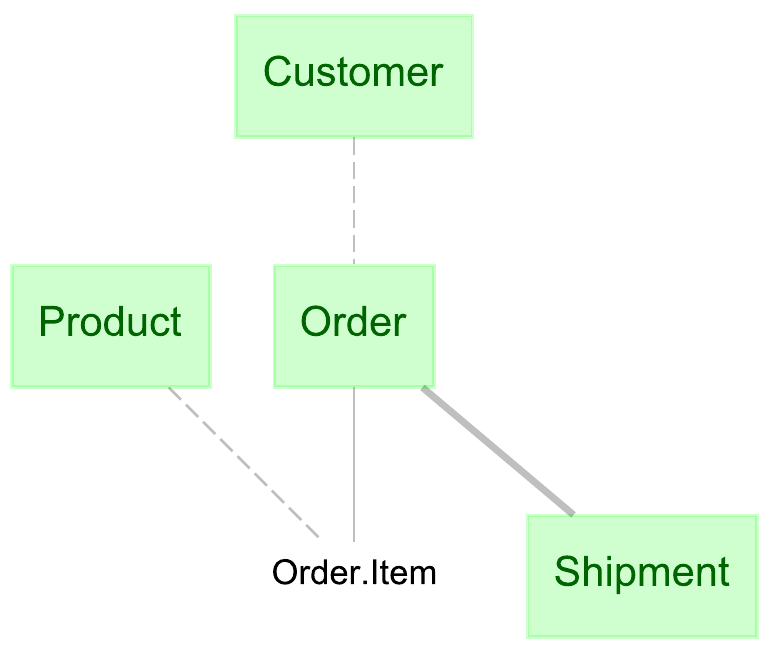}
\caption{
The diagram of the schema defined in Listing \ref{lst:master}, containing a master-part relationship between \lstinline{Order} and \lstinline{Order.Item}.}
\label{fig:master}
\end{figure}

Figure \ref{fig:master} depicts the corresponding schema diagram.  
Note that in the schema diagramming notation, part entity sets are depicted as nodes with no boxes.

The master-part relationship informs the application that the master and all its parts should always appear together or not at all.  
When inserting new data, applications must start an atomic transaction when inserting into the master and commit the transaction only when all the corresponding parts have been inserted.  
Direct deletes from part entity sets are not allowed: their data may only be deleted by deleting from their master, cascading to the parts.
This process guarantees that other users of the database never see partial data and that interrupted processes never leave incomplete data. 

The master-part relationship has important implications for dependencies.  When a downstream entity set forms a dependency on a master entity set, it may also be considered depended on all its parts as well. 

In our example, every element in  \lstinline$Order$ includes multiple dependent elements in \lstinline$Order.Item$.  
While the items are inserted, the order is not made visible to other processes.  
Once all the associated items are inserted, the entire order is committed and made visible to other agents accessing the database.
After that, elements of \lstinline$Order.Item$ cannot be deleted directly.
An order with all its items can be considered as one indivisible entity. 
Any entity that is dependent on the master  becomes also dependent on all its parts.
For example, an element of \lstinline$Shipment$ is dependent on an element of  \lstinline$Order$ with all associated elements of \lstinline$Order.Item$. 
The master-part pattern is suitable for representing cohesive groupings of elements within and across entity sets.

The \datajoint model does not allow nested master-part relationships: a part in one relationship cannot be a master in another.
This constraint is another tradeoff between representational power and simplicity.
Nested master-part relationships would also require nesting transactions whereas popular database engines only support one level of transactions.

\section{Computation}
\subsection{Programming language embedding}
In this paper, the \datajoint model is presented as a stand-alone language with its own syntax. 
However, its data manipulation and  query sublanguages translate straightforwardly into bindings in any programming language with rich object-oriented programming constructs.  
Indeed, \datajoint evolved from original MATLAB and Python implementations \citep{yatsenko-datajoint-2015}.

This section  describe computations only briefly, focusing on key concepts.  
A full implementation of computed entity sets is defined in the context of the host programming language.

\subsection{Computed entity sets}
\datajoint supports a well-defined approach to computations by defining \emph{computed entity sets}. 
A computed entity set is defined using the same notation as other base entity sets as described in Section \ref{sec:def1}.
However, instead of creating its elements using the \lstinline$insert$ command, a computed entity set computes its contents automatically.

To accomplish this, the computed entity set must define a \lstinline$make$ function on its \emph{primary dependency domain}.

\subsection{Primary dependency domain}
As discussed previously, the primary dependencies and distinguishing attributes of an entity set serve as the dimensions of the entity sets.

Consider the entity set \lstinline$Computed$ defined as
\begin{lstlisting}[language=dj]
::Computed 
-> P
-> Q
r : int      # distinguishing attribute
---
value : double   # computed value
\end{lstlisting}
Then the elements of \lstinline$Computed$ may be thought to occupy the three-dimensional space with axes \lstinline$P.proj()$, \lstinline$Q.proj()$, and \lstinline$r$.

The primary dependencies define the \emph{primary dependency domain}.  
In the case of \lstinline$Computed$, the primary dependency domain is the join \lstinline$D = P.proj() * Q.proj()$, comprising the allowed values of the  primary key attributes of \lstinline$Computed$.

\subsection{The \lstinline$make$ function}
The function \lstinline$s = Computed.make(t)$ must be defined to yield the set of elements \lstinline$s$ for any element \lstinline$t$ pulled from \lstinline$D$.
If \lstinline$Computed$ does not have any distinguishing attributes, then \lstinline$s$ will contain only one element. 
With distinguishing attributes, \lstinline$s$ will include multiple elements with different values of the distinguishing attribute.
Every element of \lstinline$s$ must define all the required attributes.

Once the \lstinline$make$ function is defined, the entity set can be automatically populated, \emph{i.e.} inserted into \lstinline$Computed$.

\section{Discussion}
In the half century since the emergence of the relational model for databases, its use has been standardized by the wide adoption of SQL.
Even so, both SQL and the relational data model lack conceptual clarity, making them unwieldy for conceptual design and for complex queries.
Computer scientists have converged on teaching a two-phased approach consisting of \emph{conceptual modeling} followed by \emph{logical modeling} prior to implementation. 

\datajoint represents a relational data model that is  more conceptually refined than earlier incarnations.
The model presented in this paper constitutes a complete query language for data definition, data manipulation, and data queries.
This conceptual refinement is summarized by the concept of entity normalization (Sections \ref{sec:norm} and \ref{sec:operational}).
A single data definition language and the schema diagramming notation are sufficiently clear to effectively communicate the overall design within the development team, with management and customers. 

The data definition language provides well-defined constructs for data integrity including entity integrity (Section \ref{sec:entity}), referential integrity (Section \ref{sec:referential}, and compositional integrity (Section \ref{sec:master}).

\datajoint's conceptual refinements resulted in a minimal query language for succinct and expressive data queries. 
For reference, Section \ref{sec:sql-trans} lists equivalent SQL and \datajoint queries for examples from the main text.

The syntax of query expressions is algebraically closed and uses binary and unary operators (Section \ref{sec:expressions}), which allows straightforward embedding in most object-oriented programming languages with only minor syntactical differences from the query languages described in this paper \citep{yatsenko-datajoint-2015}.

The \datajoint model evolved through numerous refinements while in active continuous use for processing vast amounts of data from scientific experiments. 
This paper summarizes the ground principles that will drive further improvements and the current and emerging implementations of the model. 

\section{Acknowledgements}
We thank Fabian Sinz, Christopher Turner, and Jacob Reimer for their critical review and thorough feedback on this manuscript.

We conceived \datajoint in Andreas S.\ Tolias' lab in the Neuroscience Department at Baylor College of Medicine in the fall of 2009. 
Initially implemented as a thin MySQL API in MATLAB, it defined the major principles of the \datajoint model summarized here. 

Many students and postdocs in the lab as well as collaborators and early adopters have contributed to the project.
Jacob Reimer and Emmanouil Froudarakis became early adopters in Andreas Tolias' Lab and propelled development.
Alexander S.\ Ecker, Philipp Berens, Andreas Hoenselaar, and R.\ James Cotton contributed to the formulation of the overall requirements for the data model and critical reviews of \datajoint development.

Outside the Tolias lab, the first labs to adopt \datajoint (approx.\ 2010) were the labs of Athanassios G.\ Siapas at CalTech, Laura Busse and Steffen Katzner at the University of T\"ubingen.

In 2015, the Python implementation gained momentum with Edgar Y.\ Walker and Fabian Sinz joining as principal contributors. 

In 2017, DARPA awarded a small-business innovation research grant to Vathes LLC (Contract D17PC00162) to further develop and publicize the \datajoint framework. 

In June 2018, the Princeton Neuroscience Institute, under the leadership of Prof.\ Carlos Brody, began funding a project to generate a detailed \datajoint user manual.  

\begin{appendices}

\section{SQL Translations}\label{sec:sql-trans}
This section contains translations of a selection of \datajoint queries from the main text into SQL. 

\begin{lstlisting}[language=SSQL, caption={SQL translations of queries from Listing \ref{lst:uni1}.}]
CREATE TABLE Student (
   student_id int unsigned NOT NULL COMMENT "university ID",
   first_name varchar(40) NOT NULL,
   last_name varchar(40) NOT NULL,
   sex enum('F', 'M', 'U') NOT NULL, 
   date_of_birth date NOT NULL,
   home_address varchar(200) NOT NULL COMMENT "street address",
   home_city varchar(30) NOT NULL,
   home_state char(2) NOT NULL COMMENT "two-letter abbreviation",
   home_zipcode varchar(10) NOT NULL,
   home_phone varchar(14) NOT NULL, 
   PRIMARY KEY (student_id))

CREATE TABLE Department(
   dept char(6) NOT NULL COMMENT "abbreviated department name, e.g. BIOL",
   dept_name varchar(200) NOT NULL COMMENT "full department name",
   dept_address varchar(200) NOT NULL COMMENT "mailing address",
   dept_phone varchar(14),
   PRIMARY KEY(dept))

CREATE TABLE StudentMajor(
   student_id int unsigned NOT NULL COMMENT "university ID",
   dept char(6) NOT NULL COMMENT "abbreviated department name, e.g. BIOL",
   declare_date date NOT NULL COMMENT "when student declared her major",
   PRIMARY KEY (student_id),
   FOREIGN KEY (student_id) REFERENCES Student(student_id),
   FOREIGN KEY (dept) REFERENCES Department(dept))
\end{lstlisting}

\begin{lstlisting}[language=SSQL, caption={SQL translations of queries from Listing \ref{lst:res-set}}]
#  Student & Enroll
SELECT * 
FROM Student 
WHERE student_id IN (
        SELECT student_id 
        FROM Enroll)

#  Student \ Enroll
SELECT * 
FROM Student 
WHERE student_id NOT IN (
        SELECT student_id 
        FROM Enroll)
\end{lstlisting}

\begin{lstlisting}[language=SSQL, caption={SQL translations of queries from Listing \ref{lst:res-comp}}.]
#  Student &
#    (Enroll & dept == "BIOL") \
#    (Enroll & dept == "MATH")
SELECT * 
FROM Student 
WHERE student_id IN (
        SELECT student_id 
        FROM Enroll
        WHERE dept == "BIOL")
    AND student_id NOT IN (
        SELECT student_id
        FROM Enroll
        WHERE dept == "MATH")

#  Student \ (Enroll & CurrentTerm)
SELECT *
FROM Student
WHERE student_id NOT IN (
        SELECT student_id 
        FROM Enroll
        WHERE (student_id, term_year, term) IN (
            SELECT student_id, term_year, term
            FROM CurrentTerm))

#  Student & (Enroll \ StudentMajor & CurrentTerm)
SELECT *
FROM Student 
WHERE student_id IN (
        SELECT student_id 
        FROM Enroll
        WHERE (student_id, dept) NOT IN ( 
                SELECT student_id, dept 
                FROM StudentMajor)
            AND (term_year, term) IN (
                SELECT term_year, term
                FROM CurrentTerm))

#  Student & [Enroll, StudentMajor]
SELECT * 
FROM Student
WHERE student_id IN (
        SELECT student_id
        FROM Enroll)
    OR student_id IN (
        SELECT student_id
        FROM StudentMajor

\end{lstlisting}
\end{appendices}

\bibliography{DataJoint}

\begin{thebibliography}{6}
\providecommand{\natexlab}[1]{#1}
\providecommand{\url}[1]{\texttt{#1}}
\expandafter\ifx\csname urlstyle\endcsname\relax
  \providecommand{\doi}[1]{doi: #1}\else
  \providecommand{\doi}{doi: \begingroup \urlstyle{rm}\Url}\fi

\bibitem[Codd(1970)]{codd_relational_1970}
E.~F. Codd.
\newblock A {Relational} {Model} of {Data} for {Large} {Shared} {Data} {Banks}.
\newblock \emph{Commun. ACM}, 13\penalty0 (6):\penalty0 377--387, June 1970.
\newblock ISSN 0001-0782.
\newblock \doi{10.1145/362384.362685}.
\newblock URL \url{http://doi.acm.org/10.1145/362384.362685}.

\bibitem[Kent(1983)]{kent-1983-simple}
William Kent.
\newblock A simple guide to five normal forms in relational database theory.
\newblock \emph{Communications of the ACM}, 26\penalty0 (2):\penalty0 120--125,
  1983.

\bibitem[Chen(1976)]{chen_entity_1976}
Peter Pin-Shan Chen.
\newblock The entity-relationship model --- toward a unified view of data.
\newblock \emph{ACM Transactions on Database Systems (TODS)}, 1\penalty0
  (1):\penalty0 9--36, 1976.

\bibitem[Elmasri and Navathe(2015)]{elmasri-2015-fundamentals}
Ramez Elmasri and Shamkant Navathe.
\newblock \emph{Fundamentals of database systems}.
\newblock Addison-Wesley Publishing Company, 7th edition, 2015.

\bibitem[Coronel and Morris(2016)]{coronel-2016-database}
Carlos Coronel and Steven Morris.
\newblock \emph{Database systems: design, implementation, \& management}.
\newblock Cengage Learning, 2016.

\bibitem[Yatsenko et~al.(2015)Yatsenko, Reimer, Ecker, Walker, Sinz, Berens,
  Hoenselaar, Cotton, Siapas, and Tolias]{yatsenko-datajoint-2015}
Dimitri Yatsenko, Jacob Reimer, Alexander~S. Ecker, Edgar~Y. Walker, Fabian
  Sinz, Philipp Berens, Andreas Hoenselaar, Ronald~James Cotton, Athanassios~S.
  Siapas, and Andreas~S. Tolias.
\newblock {DataJoint}: managing big scientific data using {MATLAB} or {Python}.
\newblock \emph{bioRxiv}, page 031658, November 2015.
\newblock \doi{10.1101/031658}.
\newblock URL \url{http://biorxiv.org/content/early/2015/11/14/031658}.

\end{thebibliography}

\end{document}